\documentclass[12pt]{article}

\usepackage{graphicx} % needed for figures
\usepackage{eurosym}
\usepackage{hyperref}

\providecommand{\href}[2]{#2}
\newcommand{\tseurl}[1]{\href{#1}{\texttt{#1}}}
\newcommand{\tsedoi}[1]{\href{http://dx.doi.org/#1}{\texttt{#1}}}
\begin{document}

\begin{flushright}
% \textbf{Confidential}. \\ Not to be redistributed without permission of the authors.\\
 %\texttt{Imperial/TP/13/TSE/1} \\
 European Physics Letters 2013 \textbf{104} 48001\\
 doi: \tsedoi{10.1209/0295-5075/104/48001} \\
 %\eprint{cond-mat/yymmnnn} \\
 %23rd March 1998, final corrections 8th April 1999 \\
 11th November 2013 \\
 %(\texttt{Sculplexity.tex}  LaTeX-ed on \today ) \\
 %{next preprint number}\\
\end{flushright}
 \vspace*{0.5cm}
\begin{center}
{\Large\textbf{Sculplexity:
  \\ Sculptures of Complexity
  \\ using 3D printing}}\\
\vspace*{1cm}
 {\large D.S.~Reiss, J.J.~Price and \href{http://www.imperial.ac.uk/people/t.evans}{T.S.~Evans}}
 \\
 \vspace*{1cm}
 \href{http://theory.ic.ac.uk/}{Theoretical Physics group} and \href{http://www.complexity.org.uk}{Complexity \& Networks group},
 \\
 Physics Department,
 Imperial College London,
 \\
 London, SW7 2AZ, U.K.
\end{center}

%\pacs{89.75.-k}{Complex systems}
%\pacs{07.05.Tp}{Computer modeling and simulation}
%\pacs{01.55.+b}{General physics (physics education)}

\begin{abstract}
We show how to convert models of {complex systems} such as 2D cellular automata into a 3D printed object. {Our method takes into account the limitations inherent to 3D printing processes and materials. Our approach automates the greater part of this task, bypassing the use of CAD software and the need for manual design.} As a proof of concept, a physical object representing a modified forest fire model was successfully printed. Automated conversion methods similar to the ones developed here can be used to create objects for research, for demonstration and teaching, for outreach, or simply for aesthetic pleasure. {As our outputs can be touched, they may be particularly useful for those with visual disabilities.} %}
\end{abstract}

%\keywords{3D printing; additive manufacturing technologies; complexity; forest fire model}

%\preprint{Imperial/TP/13/TSE/1}

%\maketitle

\section{Introduction}

{Complexity looks at the emergence of large scale structure from microscopic rules, usually in systems which are out of equilibrium \cite{CM05}. Over the last few decades such a complex systems viewpoint has influenced physics in many ways. For instance, this can be seen in the idea that some systems can have structure on all scales, fractals \cite{B96a}, or that the dynamics of some systems drives them to sit at critical points where large fluctuations exist, Self-Organised Criticality \cite{J98}. The emergence of so many fat-tailed (power-law like) distributions in network data can be understood through simple generic local processes \cite{ES05}. A complex systems approach has also become a valuable viewpoint for many other disciplines in physical, biological and social sciences \cite{CM05,B96a,W84,J98,ES05,B04,LPLW09}.}

\begin{figure}[ht]
			\centering
			\includegraphics[width = 16cm]{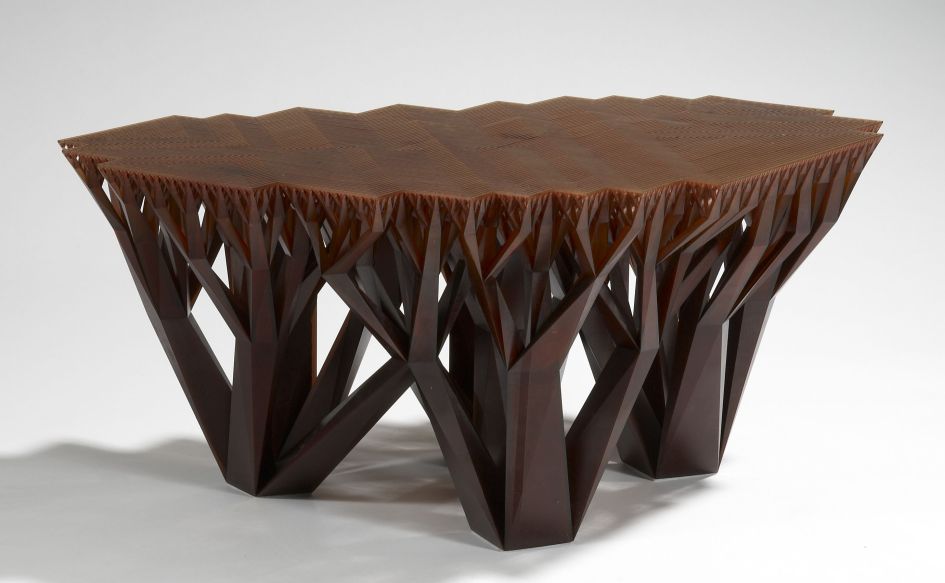}
			\caption{The Fractal.MGX coffee table, designed by WertelOberfell in collaboration with Matthias B\"{a}r \cite{Fractal.MGX}. Created out of epoxy resin using a stereo-lithography 3D printing process. (Image: WerterOberfell GbR and Stephane Briolant).}
\label{ffractalmgx}
		\end{figure}

Discussions of complex systems are often couched in the language of mathematics. This can hinder the use of complexity ideas outside physical science disciplines and makes outreach to the wider public more difficult. It is therefore important to find less technical representations of these ideas. There are some beautiful 2D visualisations, e.g.\ of fractals \cite{B96a}, but so far nothing has exploited the potential of 3D printing. As the US President Obama stated in his 2013 State of the Union address, ``new workers are mastering the 3D printing that has the potential to revolutionise the way we make almost everything'' \cite{O13}. Now that 3D printing is an affordable technology and is becoming widely available, can we use it to provide new visualisations of complex systems? Creating printable physics content may be instructive for researchers and for students in the classroom at school or university. The objects themselves can connect the wider public to current research ideas. After all unlike 2D pictures or videos, 3D printed objects can be touched making them useful for those with poor sight. Alternatively we may create these objects simply for pleasure. This project was inspired by the Fractal.MGX table \cite{Fractal.MGX} shown in Figure~\ref{ffractalmgx}, the first 3D printed object acquired by the Victoria and Albert museum in London. This table is both a beautiful object and a representation of a branching process \cite{AB04}.

{In this letter we ask if 3D printing can capture some of the fascinating ideas contained under the umbrella of ``Complexity Science''. What are the problems and how can we overcome them in making this translation? Could this be packaged in a user friendly manner? Can we create sculptures of complexity?}

\section{3D printing technologies}\label{sec:printprocesses}

{There are four common types of 3D printing \cite{GRS10}. Also known as additive manufacturing technologies, these are processes in which three-dimensional objects are formed through application of successive layers of material.}

\begin{figure}[ht]
			\centering
			\includegraphics[width = 6cm]{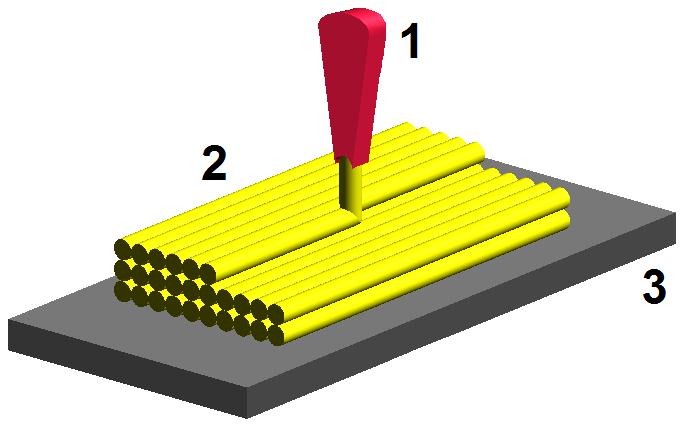}
			\caption{The Fused Deposition Modelling (FDM) extrusion process. 1 - Nozzle extruding molten plastic. 2 - Layers of material. 3 - Heated bed. (Image adapted from \texttt{https://en.wikipedia.org/wiki/File:FDM\_by\_Zureks.png}).}
			\label{ffdm}
		\end{figure}

{In Extrusion Deposition methods, melted thermoplastic is deposited in layers. Typically a heated moveable bed is placed underneath a heated nozzle which extrudes molten plastic onto the bed. Directions from a computer tell the nozzle how much to extrude and how the bed and nozzle should move with respect to each other in order to produce the desired part. The most common form of extrusion deposition is called Fused Deposition Modelling (FDM), as depicted in Figure~\ref{ffdm}. FDM is popular as it is very affordable, it is easy to install, and it can create quite complex geometries. This is the method we used to produce our final output. The major disadvantage of FDM is that printing large overhangs is often not possible \cite{GRS10}, a problem we will discuss in detail below.}

{A second approach is known as Granular Materials Binding. The most common form is Selective Laser Sintering (SLS) in which a laser is used to fuse granules of an appropriate material. As unfused granules can support the printed structure, SLS can produce structures with a high geometric complexity and significant overhangs \cite{GRS10}. Laminated Object Manufacturing (LOM) is the third method. It involves sheets of material laid layer upon layer, with a laser removing the unwanted material from each layer. The popularity of LOM is now waning with the emergence of FDM and SLS \cite{GRS10}. Finally we have Photo-polymerisation techniques in which photo-polymers are exposed to UV light causing the material to become solid. One example of this process is known as Stereo-lithography (SLA). SLA was used to build the Fractal.MGX table
\cite{Fractal.MGX} of Figure~\ref{ffractalmgx} out of epoxy resin. SLA can be used with a wide variety of materials. For instance, the process is used for bio-printing which aims to 3D print organs and tissue. However, SLA remains expensive \cite{GRS10}.}

\subsection{Constraints}

\begin{figure}[ht]
			\centering
			\includegraphics[width = 7.67cm]{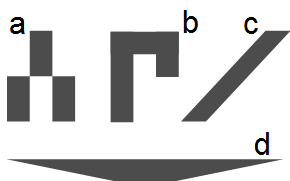}
			\caption{{Some structures which violate the constraints imposed by many (but not all) 3D printers, with gravity pulling from top to bottom. Structure (a) is connected only by edges, failing the support constraint. Structure (b) has a part with no support from below and will also violate the support constraint. Example (c) will simply topple over during printing, a violation of the stability constraint. Shape (d) has an overhang which is probably too steep, breaking the overhang constraint.}}
			\label{fig:3dconstraints}
		\end{figure}

To create a physical object using 3D printing technologies, we must consider the limitations of the printing process. Since these limitations are not always very precise, we {focus} on the key geometric constraints of the {mass market} 3D printing processes available, though these limitations are common to most (but not all) 3D printing processes. There are three main geometrical limitations (see {Figure}~\ref{fig:3dconstraints}):
\begin{description}
		\item[Support:] Each section of the geometry must be {sufficiently} supported from below by another section of the {shape}. The foundation layer is supported by the bed of the 3D printer.
		\item[Overhang:] The surface cannot be at too steep {an} angle to the vertical. If higher layers have too much of an overhang then the structure may deform, droop or warp during the printing process. {A safe limit for most 3D printing processes is that surfaces should be at no more than $30^\circ$ to the vertical.}
		\item[Stability:] The geometry must be such that at any point in the printing process the object remains static and does not {topple} under its own weight.
\end{description}
We will show {how to deal} with these constraints {as we outline} our conversion process below.

\subsection{Input} {To print an object, 3D printers take a file describing the surface of the object to be printed. Generally this is a list of triangles describing an approximation to the desired shape. To produce these files, a designer normally uses CAD (computer aided design) software, some of which is free.} However, the key idea in our approach is that we can use the height of these 3D printed objects to represent time in a theoretical model. Each layer of a 3D printed object is then matched with two spatial dimensions in a complex system model. Our aim is to produce a file specifying a triangulated surface automatically, only requiring parameter values and the initial conditions for a complex system model.

\section{Complex Processes}

{Many models of complex systems are examples of cellular automata} \cite{W84,J98,CM05}. Cellular automata may be defined as a collection of identical cells, typically arranged in a lattice, each of which is in exactly one of a finite number of states. The cells evolve synchronously in discrete time, changing state according to local interactions and a given set of rules.

To illustrate our approach we shall use a {two-dimensional stochastic cellular automaton} on a square lattice, a modified version of the forest fire model \cite{J98,DS92}. In our system, each cell on a square lattice may be in one of three states: alive, dead, or burning. The rules which dictate the evolution of the system are given as follows:
\begin{description}
	\item[Rule 1:]{A burning cell always dies.}
	\item[Rule 2:]{An alive cell starts burning with probability~$q$ if one or more of its neighbours are burning, as if a fire were spreading.}
	\item[Rule 3:]{A dead cell which is adjacent to an alive cell becomes alive with probability~$p$, as if a forest were spreading.}
	\item[Rule 4:]{An alive cell with no burning neighbours starts burning with probability~$f$, as if struck by lightning.}
\end{description}

\begin{figure}[ht]
			\centering
			\includegraphics[width=7.98cm]{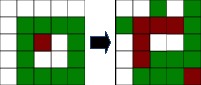}
			\caption{An illustration of the modified forest fire model for one time step. The green (light) shaded squares represent trees which are alive, red (dark) squares are burning trees, and white squares are empty. The single burning tree on the left disappears (rule 1) but it's flames spread to all of its neighbours which were alive (rule 2 with $q=1$). One of the trees which was previously alive, in the bottom right hand corner, is hit by {lightning} (rule 4, typical for $f \sim 0.14$). Meanwhile {five} new trees have sprouted in the right hand picture (rule 3, typical for $p \sim 0.41$).}
			\label{fmffm}
		\end{figure}

\section{Conversion Process}

We start from a theoretical model which defines a series of points, $(x,y,t)$, which are either ``occupied'' or ``empty''. For example, for our modified forest fire model we treat any cell with an alive tree at time $t$ and at lattice point $(x,y)$ as occupied, otherwise points are treated as empty. As we use a square lattice in our forest fire example, we will take the coordinates to be integers, with a spacing of one in both the time and space dimensions. Our first task is to convert these occupied points in the theoretical model space into a series of shapes in the three-dimensional space used by the 3D printer with coordinates $(X,Y,Z)$. We assume gravity acts in the negative $Z$ direction and that $X$ and $Y$ are orthogonal horizontal coordinates.

\begin{figure}[ht]
	\centering
	\includegraphics[width = 16cm]{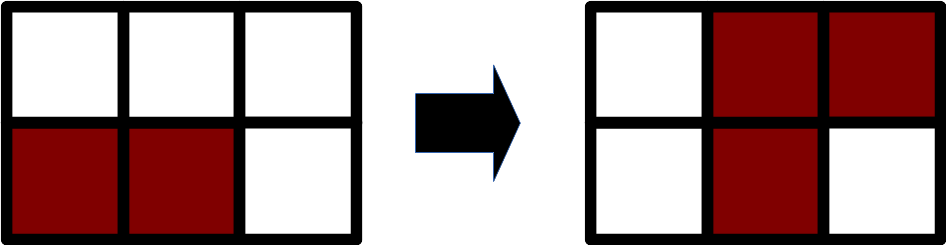}
	\caption{The evolution in a theoretical model is typically shown as a sequence of two-dimensional lattices --- here time increases from left to right. We convert these into a three-dimensional shape by treating each time slice as a layer of cubes and then stacking these in temporal order as in Figure~\ref{fig:slabstack}.}
	\label{fig:beforeextrude}
\end{figure}

\begin{figure}[ht]
	\centering
	\includegraphics[width = 16cm]{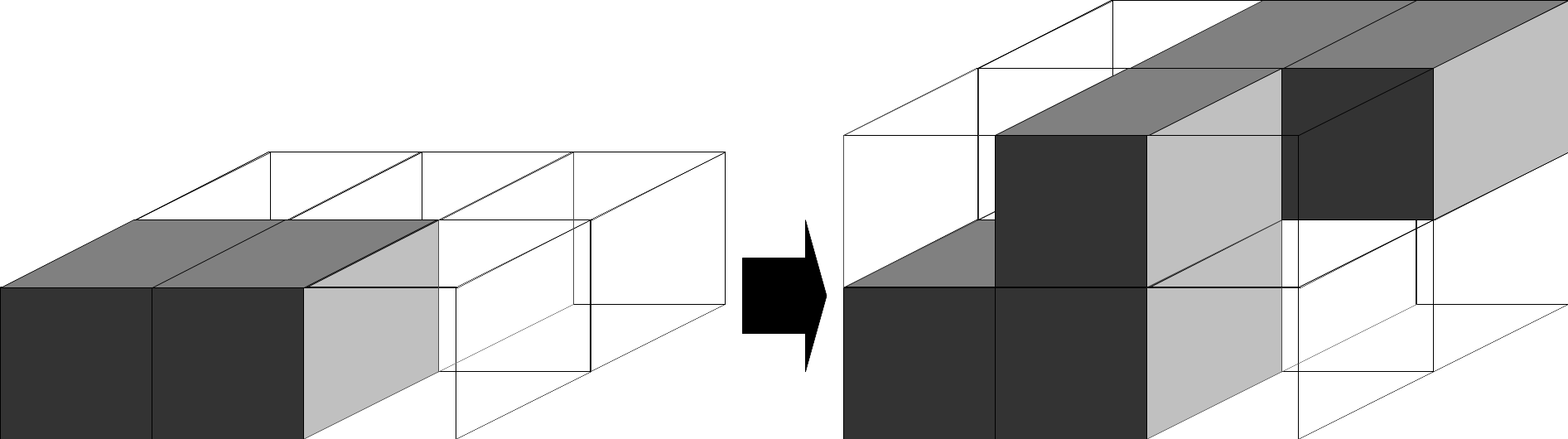} %EPL
	\caption{{The {simplest} representation of the evolution of the theoretical model of Figure~\ref{fig:beforeextrude}. At each time, the two-dimensional lattice becomes a 3D layer of cubes. We then stack successive time layers together to establish a three-dimensional object. Here time increases as you move up the page, such that the layer representing the subsequent time step is placed on top of the slab shown on the left.}}
	\label{fig:slabstack}
\end{figure}

{The basic idea is to envisage our binary two-dimensional lattices at each time $t$, such as those in Figure~\ref{fig:beforeextrude}, as a single layer of 3D cubes with centres at height $Z$ as shown in Figure~\ref{fig:slabstack}. Each occupied site (alive trees in our case) at point $(x,y,t)$ is represented by a filled cube with corners at $(X,Y,Z) = (x \pm \frac{1}{2} ,y\pm \frac{1}{2} , t\pm \frac{1}{2} )$. These layers are then simply stacked upon one another, creating a 3D array of cubes, as illustrated in Figure~\ref{fig:slabstack}.}

{For the final step we need to detect the outer surfaces of our cubes and represent these surfaces in terms of triangles. It is this list of triangles which is used as the input to 3D printers. However, we have not addressed the constraints 3D printers impose upon their output and so we now turn to this difficult aspect.}

\subsection{Constraints and Process Refinement}

In designing a conversion process, the main factors which control the geometry of the final object are:
\begin{itemize}
		\item{The choice of theoretical model to be converted. This includes the lattice type, model parameters, and update rules.}
		\item{The initial conditions and {the number of time steps used in the evolution of the theoretical model}.}		
		\item{The method used to convert the set of theoretical model points $(x,y,t)$ into a 3D object. This is how the geometry is introduced and there is much freedom in the choice of this method.}
\end{itemize}

The issue of support immediately suggests that we can not allow spontaneous creation of occupied sites in our theoretical model as this would correspond to material hanging in mid-air in our final 3D object. This is an example of how theoretical model choices are important in satisfying the constraints. {For example the standard forest fire model \cite{J98,DS92} uses a different rule 3 as it allows \emph{any} empty cell to become filled with probability $p$. This can create isolated new `trees' which correspond to unsupported floating components in the final 3D shape. Such unsupported parts are not printable. Our version of rule 3 was chosen to ensure that the parts of the 3D object representing new growth can be supported from the parts representing existing trees.}

Stability problems can be dealt with through initial conditions. In the case of our modified forest fire model, we {choose} to start with three or more patches of alive trees which will go on to form `legs' for our final 3D printed object. As any patch of trees could die out and provide no support, we still have to check each individual run of our theoretical physics model. {We find it is simple to do this by hand, though analysing mechanical stability of the final object could in principle be automated.}

\begin{figure}[ht]
	\centering
	\includegraphics[width = 16cm]{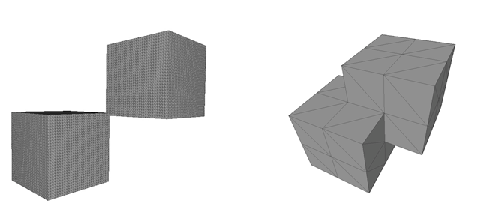} %EPL
	\caption{The simple solution of stacking layers to create a 3D array of cubes will often create situations where {some cubes touch only along edges or at corners, as shown on the left. This can leave some cubes without sufficient mechanical support. By associating larger cubes with each occupied point in the theoretical model we can address most of these support issues.}}
	\label{fig:nooverlap}
\end{figure}

{The support constraint is more difficult to deal with.
The simple mapping suggested above, points $(x,y,t)$ becoming cubes with corners at $(X,Y,Z) = (x \pm \frac{1}{2} ,y\pm \frac{1}{2} , t\pm \frac{1}{2} )$, can leave us} with cubes connected by a single corner or edge, as seen in {Figures~\ref{fig:3dconstraints}a and \ref{fig:nooverlap}.} These are clearly structurally unsound. To fix this problem, we {expand} the size of each cube so that neighbouring cubes are overlapping, {as shown in Figure}~\ref{fig:nooverlap}. Thus an occupied cell in the theoretical model at $(x,y,t)$ becomes a cube in the 3D object with corners at $(X,Y,Z) = (x \pm 1,y \pm 1,t \pm 1)$. This mapping {ensures} all cubes have support from at least one neighbour. {Mechanical strength and stability on larger scales is still not guaranteed but we find that visual inspection and repeating runs of our stochastic model are sufficient to deal with any remaining issues of support.}

The 3D array of overlapping cubes still has a problem with overhangs. This happens both at the small scale, when looking at the 90$^\circ$ angles between surfaces of overlapping cube-cube connections, and on the {scale of several time steps}.

Let us consider first the large scale overhangs. If we imagine an extended series of cubes of the type shown in {Figure}~\ref{fig:nooverlap}, then we can create large scale surfaces at 45$^\circ$ to the vertical, {whereas 30$^\circ$ is a safe limit for most 3D printers.} Our solution {is} to stretch the mapping of time $t$ in the theoretical model to the height $Z$ of the 3D object. Thus we map occupied points in the theoretical model $(x,y,t)$ to cuboids with corners $(X,Y,Z) = (x \pm 1,y \pm 1,v(t \pm 1) )$ {where a very safe value for $v$ would be} $\cot(30^\circ) \approx 1.73$.

Finally we need to convert our series of overlapping cuboids into a single object and, further, one where we can define its surfaces in terms of triangles. This task is greatly simplified because we are working in terms of a square lattice in the theoretical model space and cuboids in the 3D object space. First we split each of our cuboids $(X,Y,Z) = (x \pm 1,y \pm 1,v(t \pm 1) )$ into eight equal sized cuboids, all sharing a common point at the centre $(X,Y,Z) = (x,y,vt )$. {These smaller cuboids} either overlap perfectly (and we drop such duplicates easily), or not at all. This avoids any need to analyse the partial overlap of 3D shapes, {a non-trivial exercise}. Working with these smaller non-overlapping cuboids means it is also easy to {determine} if any of their surfaces are external and thus part of the surface of the whole 3D object. We need only check if two cuboids are adjacent (their centres differ by one unit in one coordinate, and are identical in the others) to realise that they share a common face {which is therefore} an internal surface. Once we {know} which of the rectangular faces of the small cuboids are external, it is simple to represent this face as two triangles. This is shown in {Figure}~\ref{fig:nooverlap}. Defining an object in terms of its external surface using a series of triangles is a standard method and there are widely used, simple file formats which do this, such as STL (Standard Tessellation Language) \cite{STL}. We are now able to pass our three-dimensional object representing a complexity model to standard 3D packages.

We are still left {with overhangs at the smallest defined scale.} The model, as it is now defined, has unprintable horizontal surfaces on the bottom face of any overhanging cube, as seen in {Figure}~\ref{fig:nooverlap}. Luckily, typical software used to manipulate 3D objects have smoothing filters which are effective on these small scales. {Such filters remove all sharp edges and create smooth surfaces in place of our 90$^\circ$ overhangs. We use the smoothing filter `Laplacian Smooth' in the mesh editor MeshLab \cite{MeshLab}, and found this sufficient for our purposes.}

\section{Results}

\begin{figure}[ht]
	\centering
	\includegraphics[width = 16cm]{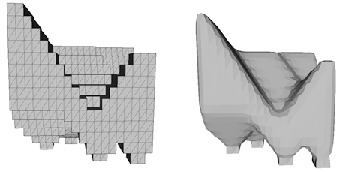} %EPL
	\caption{Rough and smooth representations of the modified forest fire model. A boundary of fixed empty cells is used.}
	\label{fig:fireexpand}
\end{figure}

\begin{figure}[ht]
	\centering
	\includegraphics[width = 16cm]{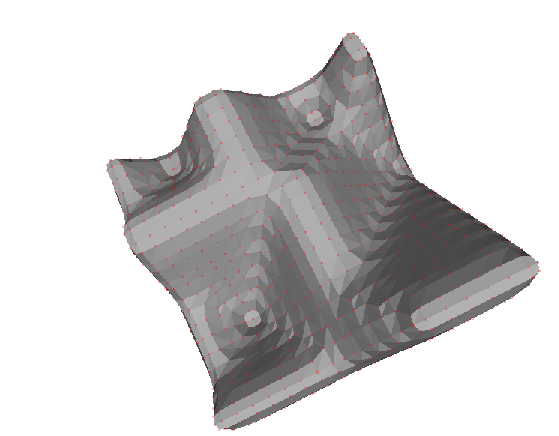}
	\caption{A top down view of a modified forest fire model, highlighting the `pits' caused by lightning strikes.}
	\label{fig:firepits}
\end{figure}

The complete story of the evolution of our modified forest fire model can be extracted from the 3D geometry. In {Figure}~\ref{fig:fireexpand}, you can see the expansion of `forests' from several individual `trees' at the base, which eventually merge into each other. The pit-like formations, better seen in {Figure}~\ref{fig:firepits}, show where and when `lightning' struck a `tree'.

The smoothed 3D model depicted in {Figure}~\ref{fig:fireexpand} was chosen to be 3D printed as a proof of concept of the process and software we developed \cite{Reiss,Price,SculplexityCode}. We used a RepRap v2.0 \emph{Mendel} 3D printer\ (currently costing around {\euro 650} in parts and must be self-assembled), which uses a Fused Deposition Modelling printing process to print the 3D model in black ABS plastic. The threshold overhang angle for this process and material is typically around 45$^\circ$, and so the z-dimension was stretched by a factor of 1.5 to allow for a margin of error. The printing took around 8 hours and used around {\euro 15} of material.

\begin{figure}[ht]
	\centering
	\includegraphics[width = 7.5cm]{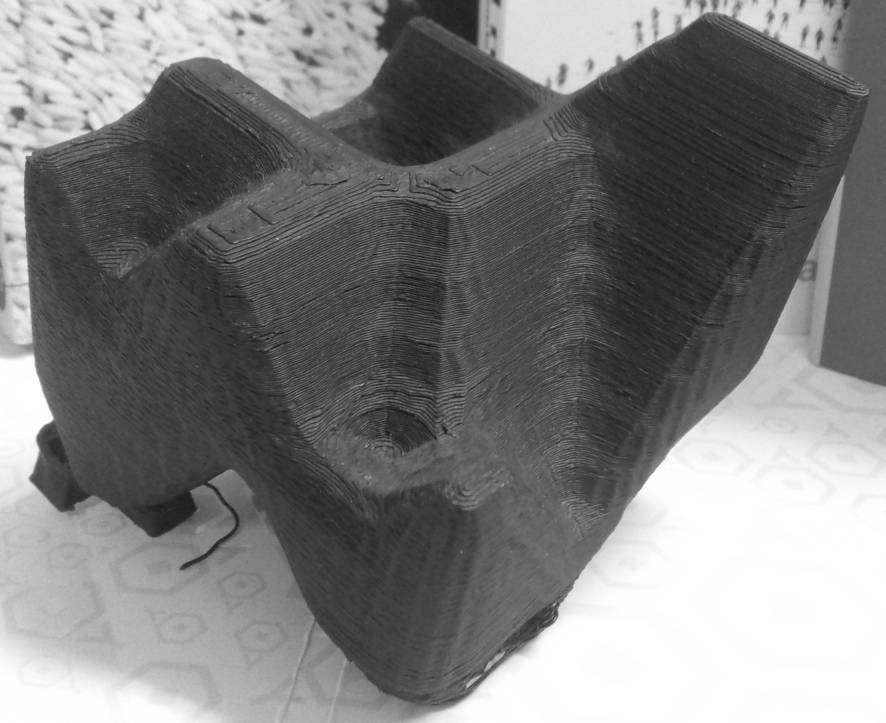}
	\includegraphics[width = 7.5cm]{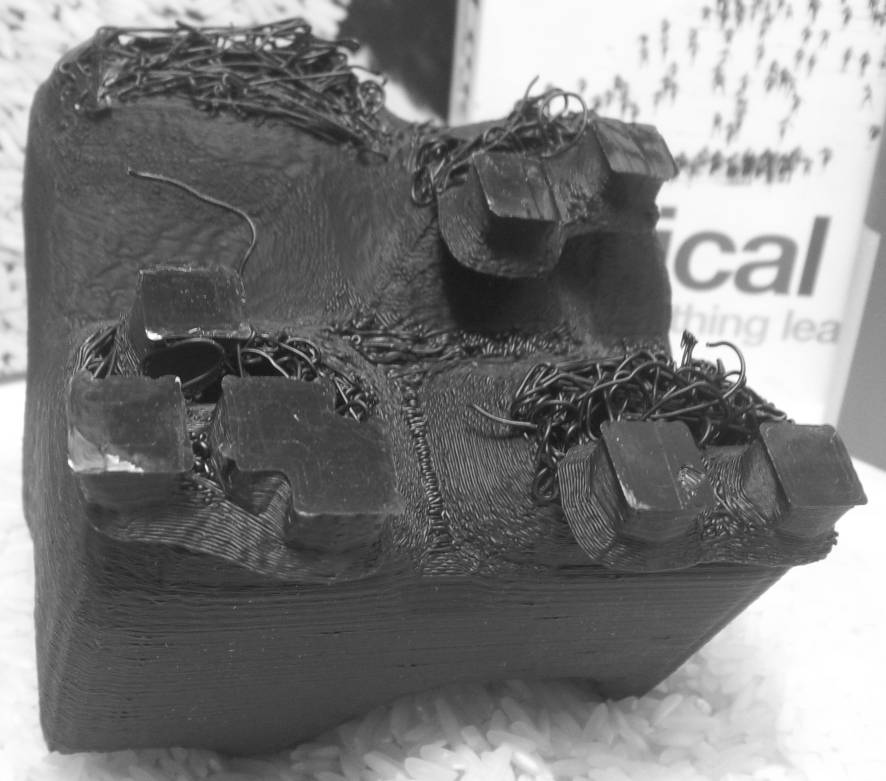}
	\caption{Two views of the final printed modified forest fire model depicted in Figures~\ref{fig:fireexpand} and~\ref{fig:firepits}. This 3D printed object would just fit into an (8cm)$^3$ cube. While some issues can be seen at the base, for the most part this model shows that the process developed throughout the project is successful in transforming a model of complexity, a 2D cellular {automaton}, into a 3D printable geometry.}
	\label{fig:3dmodel}
\end{figure}

The 3D printing process was relatively successful, producing the physical 3D model depicted in {Figure}~\ref{fig:3dmodel}. However, some issues were encountered at the beginning of printing, affecting the bottom of the model. The thin string-like structures visible at the base are the result of the Fused Deposition Modelling process attempting to print with no supporting structure, caused by an unintended offset of the upper-section relative to the base. The reason for this may have been that some of the structures in the base of our model were too thin. Thin structures may curl upwards significantly when cooling, putting them in the way of the unknowing extrusion nozzle. {This type of problem is common with 3D printing. It is often fixed by simply adjusting some of the printer's settings, such as the temperatures of the bed and extrusion head or the speed of the extrusion nozzle}. {This} illustrates that 3D printing is currently as much a craft as it is a science.

\section{Conclusions and future work}

We have shown how a model of complexity, a 2D cellular automaton, can be translated into a 3D object using current inexpensive 3D printing technology. We used the height (time of deposition) of each layer in the printed object to represent time in the theoretical model, leaving the other two dimensions in the 3D object to represent the spatial dimensions of the theoretical model. There are considerable obstacles in making this translation which arise from the limitations of typical 3D printers. However, we have identified these issues and shown how to deal with all of them. Finally, using low cost readily available technology, we produced a three-dimensional object which represents one run of a modified forest fire model as a proof of concept. All code written and used in the project can be found online \cite{SculplexityCode}.

Despite the success of our approach, there is still much work to be done in this area. While we have identified one approach, there are many other ways of making the translation from theoretical model to 3D object.
In terms of applications, we have not fully investigated the parameter space of the modified forest fire model featured here. We also tried some lattice gas models \cite{J98,J90} but while we could produce printable objects, {they seemed relatively uninteresting} \cite{Reiss,Price}. {There are of course many other models worth investigating. For researchers studying models of complexity or cellular automata, a 3D realisation might {reveal} features which were not obvious from traditional 2D figures.} Classifying different emergent 3D features by the parameter space could be useful and interesting physically.
In terms of outreach, both to physics students or the wider public, we have yet to see how such objects can be used though informal feedback has been very positive. {As people may touch and handle our outputs, they may be particularly useful for those with impaired sight.}
Finally our result is not particularly aesthetically pleasing. It would be interesting to think how to adapt our framework to produce objects of beauty in their own right, much as the Fractal.MGX table \cite{Fractal.MGX} of Figure~\ref{ffractalmgx} represents both a branching process \cite{AB04} {and an object of aesthetic value}.

Our work also highlights a lack of freely available resources documenting the quantitative limitations of specific 3D printing processes and materials. There exist tools which check the validity of STL files, looking for syntactical errors or non-manifoldness. However, no utility exists to check if a STL file is actually 3D printable for a given process/material. Such a tool would be very helpful and would reduce costs, especially for non-experts of which there are bound to be many more due to the growth of 3D printing.

Hopefully the physics community can contribute to further developments. {There is a great opportunity here for new representations and visualisations as our results illustrate. There already exist some inexpensive printers can print several colours at once, uncovering new potential innovations. For instance, colour could be used to represent the different states of a cellular automaton. The technology continues to evolve rapidly.} In any case, the physics community can only benefit by taking part in this technological revolution.

Finally it is interesting to note that ours is not the first attempt to convert theoretical models from complexity into physical three-dimensional objects. Seventeen years ago Rage et al.\ \cite{RFWWCS96} produced a three spatial-dimensional shape (fixed in time) defined using a Diffusion Limited Aggregation process. However, their approach used plans from a Fortran programme, balsa wood, and glue. This highlights how our methods, using 3D printing and C++ programmes with a GUI interface, have now moved this idea forward.

\section*{Acknowledgments}
We would like to thank Thomas Clayson and Aksat Shah for their expert help with 3D printing and the Imperial College Robotics society for our use of their 3D printer. We also thank Doug Plato for his comments on the manuscript.

%% *****************************************************************
%%
%% BIBLIOGRAPHY
%%
%% **************************************************************
%% Uses BibTeX
%%
%
%%\newpage
%\bibliographystyle{unsrt}
%%\bibliographystyle{abbrv}
%%\bibliography{/DATA/JabRef/TSEjabrefdatabase}
%\bibliography{/PAPERS/JabRef/TSEjabrefdatabase}

%\section*{References}

\newcommand{\Name}[1]{#1}
\newcommand{\Book}[1]{\emph{#1}}
\newcommand{\Publ}[1]{#1}
\newcommand{\Year}[1]{(#1)}
\newcommand{\REVIEW}[4]{#1 \textbf{#2} (#3) #4}
\newcommand{\papertitle}[1]{``#1''}

\end{document}